\documentclass[conference,twocolumn]{IEEEtran}
\IEEEoverridecommandlockouts
\usepackage{cite}
\usepackage{amsmath,amssymb,amsfonts}
\usepackage{algorithmic}
\usepackage{graphicx}
\usepackage{textcomp}
\usepackage{xcolor}
\usepackage{tensor}
\usepackage{caption}
\usepackage{bm}
\usepackage{float}
\usepackage{subfig}
\usepackage{booktabs}

\def\BibTeX{{\rm B\kern-.05em{\sc i\kern-.025em b}\kern-.08em
    T\kern-.1667em\lower.7ex\hbox{E}\kern-.125emX}}

\begin{document}

\title{\huge Optical Integrated Sensing and Communication for Cooperative Mobile Robotics: Design and Experiments}
\author{\IEEEauthorblockN{Shengqian Wang\textsuperscript{1} and He (Henry) Chen\textsuperscript{1,\,2}}
\thanks{This research was supported in part by project \#MMT 79/22 of the Shun Hing Institute of Advanced Engineering, The Chinese University of Hong Kong. The work of S. Wang was supported by the Hong Kong PhD Fellowship Scheme (PF20-48158). 

Email:{\{ws021, he.chen\}@ie.cuhk.edu.hk}
}
\IEEEauthorblockA{\textsuperscript{1}Department of Information Engineering, The Chinese University of Hong Kong, Hong Kong SAR, China} 
\IEEEauthorblockA{\textsuperscript{2}Shun Hing Institute of Advanced Engineering, The Chinese University of Hong Kong, Hong Kong SAR, China} 
}

\maketitle

\begin{abstract}
Integrated Sensing and Communication (ISAC) is an emerging technology that integrates wireless sensing and communication into a single system, transforming many applications, including cooperative mobile robotics. However, in scenarios where radio communications are unavailable, alternative approaches are needed. In this paper, we propose a new optical ISAC (OISAC) scheme for cooperative mobile robots by integrating camera sensing and screen-camera communication (SCC). Unlike previous throughput-oriented SCC designs that work with stationary SCC links, our OISAC scheme is designed for real-time control of mobile robots. It addresses new problems such as image blur and long image display delay. As a case study, we consider the leader-follower formation control problem, an essential part of cooperative mobile robotics. The proposed OISAC scheme enables the follower robot to simultaneously acquire the information shared by the leader and sense the relative pose to the leader using only RGB images captured by its onboard camera. We then design a new control law that can leverage all the information acquired by the camera to achieve stable and accurate formations. We design and conduct real-world experiments involving uniform and nonuniform motions to evaluate the proposed system and demonstrate the advantages of applying OISAC over a benchmark approach that uses extended Kalman filtering (EKF) to estimate the leader's states. Our results show that the proposed OISAC-augmented leader-follower formation system achieves better performance in terms of accuracy, stability, and robustness.
\end{abstract}
\section{Introduction}
In recent years, cooperative mobile robotics has played an increasingly important role in a wide range of industrial applications, including cooperative transportation \cite{machado2016multi}, exploration \cite{yasuda2012adaptive}, and surveillance \cite{khaleghi2014dddams}. This technology offers promising performance, high efficiency, and effectiveness for operations that are dangerous or labor-intensive for humans \cite{lins2021cooperative, ronzoni2021support, lytridis2021overview}. Information perception and sharing using onboard sensors and wireless communications are crucial for cooperative robots to complete tasks, but traditionally, sensing and communications have been performed separately. However, the development of massive multi-input multi-output (MIMO) and millimeter wave (mmWave)/terahertz (THz) technologies has enabled the integration of sensing and wireless communications in shared hardware and spectrum resources \cite{cui2021integrating, liu2022integrated, liu2022survey, tan2021integrated}. Integrated Sensing and Communication (ISAC) offers new services and potential solutions for cooperative mobile robotics to address bottleneck challenges such as high-accuracy localization and tracking \cite{tan2021integrated} and achieve better performance.
\begin{figure} 
	\centering
          \subfloat[\label{fig: robot setup}]{
		 \includegraphics[scale=0.11]{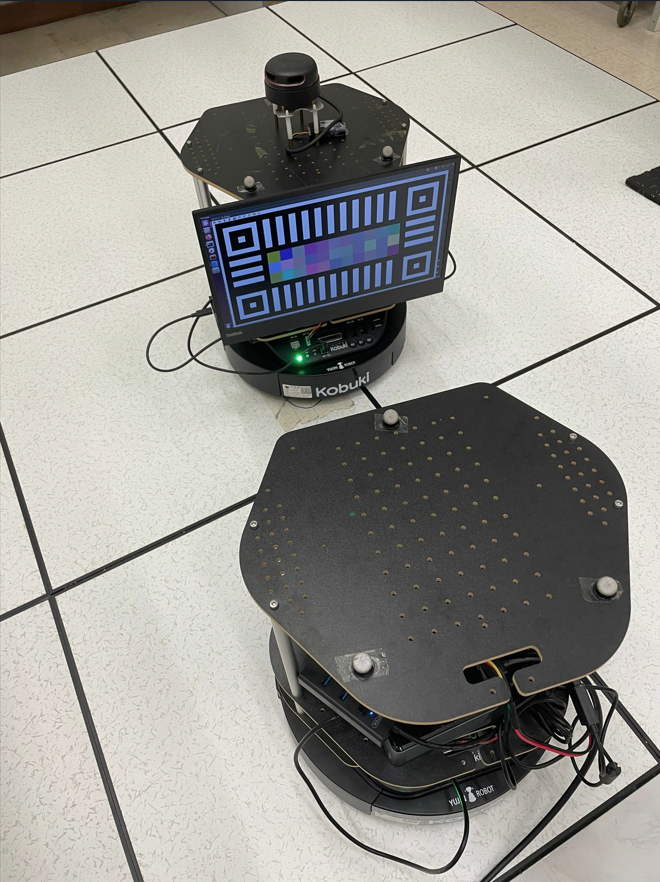}}
	   \subfloat[\label{fig: occ}]{
		 \includegraphics[scale=0.11]{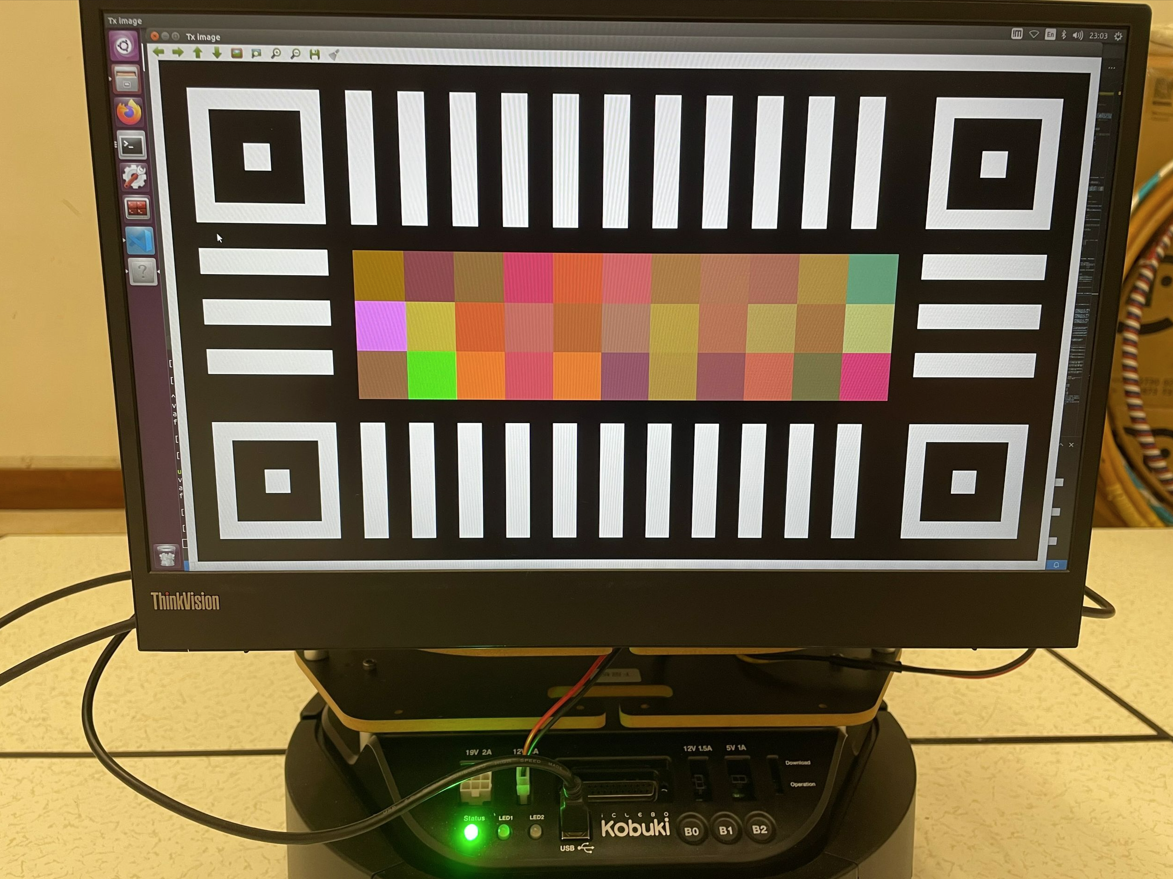}} 
	\caption{Experimental setup. (a) Two Turtlebot2. (b) An illustrative feature image displayed on a 15.6-inch LCD screen.}
	\label{fig: equipment setup} 
	\vspace{-1.5em}
\end{figure}

Reliable radio communications have been the mainstay of current ISAC designs. However, radio-based solutions can be severely impacted in the presence of radio interference, radio attacks or in scenarios unsuitable for radio propagation, such as underwater environments \cite{li2014survey} and some military environments \cite{lin2018communication}. In such cases, radio-based solutions can have severely degraded performance or even lose functionality. As a result, many cooperative mobile robotics researchers are shifting their focus towards designing radio-free or sensing-only approaches to eliminate the dependence on radio communications \cite{bautin2011towards, kalat2018decentralized, cai2021behavior}. Low-cost cameras are a preferable alternative to expensive sensors such as LiDAR, and camera sensing has been widely used in various cooperative mobile robotics applications supported by well-developed computer vision technologies \cite{giusti2012cooperative, khan2016cooperative, wang2016adaptive}. However, the absence of communication inevitably leads to a decline in performance and application limitations of cooperative robot systems, despite tremendous efforts being spent on advanced and complicated sensing technologies to make up for it. In this context, inspired by radio ISAC, it is natural to ask \textit{whether it is possible to design an optical ISAC (OISAC) scheme for cooperative mobile robotics that integrates camera sensing and optical communications in a single optical channel}. This would enable robots to simultaneously perceive and receive information using only low-cost cameras, maximizing the usage of optical channels. Such a system could help overcome the limitations of radio communications and provide a more reliable and cost-effective solution for cooperative mobile robotics.

As an initial effort to answer the question above, this paper develops a new robot operating system (ROS)-compatible OISAC scheme for cooperative mobile robotics that seamlessly integrates camera sensing and screen-camera communication (SCC). Our choice to make the OISAC scheme compatible with ROS was motivated by the fact that ROS has become the de facto software platform for robot design and programming. We choose the vision-based leader-follower formation control problem as our case study to evaluate the gain of the proposed OISAC scheme, since it represents a typical problem in cooperative mobile robotics. Most recent vision-based leader-follower formation control research has been based on camera sensing technologies, such as extended Kalman filtering (EKF) and high-gain observer-based output feedback algorithms that use observation of particular markers fixed on the leader to estimate its states, see e.g., \cite{mariottini2007leader, tron2016distributed, moshtagh2009vision} and references therein. However, we realized that because of no information communication, previous camera sensing-only methods may make the follower less agile to the dynamic changes of the leader's movements, which has been confirmed by experiments presented in Sec.~\ref{section: experiments}. On the other hand, SCC is a form of optical camera communication technology that uses a screen to display images with data encoded in certain visual patterns, and a camera to capture the images and subsequently decode the embedded data \cite{perli2010pixnet, hu2013lightsync, wang2015inframe++}. We propose to integrate camera sensing and SCC so that both sensing and communication can be realized over one optical link and thus the follower can become more responsive to the changes of the leader’s movements. Nevertheless, existing SCC frameworks designed for static optical links become no longer applicable in the considered application involving robot movements and delay-sensitive control. 


The main contributions of this paper are three-fold: \textit{Firstly}, we develop a new ROS-compatible OISAC scheme using only RGB images, in which a commodity LCD screen is mounted on the leader robot to display visual information, while a low-cost camera is fixed onboard the follower robot to capture RGB images. We carefully design the displayed images on the screen to realize sensing and communication through the same optical channel. We also propose new algorithms to combat the image blur caused by robot shaking and reduce the long image display delay.  
\textit{Secondly}, we design and implement a new vision-based leader-follower formation system using our OISAC scheme, which enables the follower to simultaneously receive the leader's latest velocity and sense its relative pose to the leader. To leverage all the information acquired by the camera to achieve
stable and accurate formation, we devise a new leader-follower formation controller with low computational complexity. The stability of our controller is proven by the Lyapunov stability theory \cite{sastry2013nonlinear}. Our OISAC-based design enables the follower to react more agilely to the leader's dynamic movements. This is because directly reading out the leader's latest velocity from the captured images is far more accurate and up-to-date than any estimation, eliminating the need for complicated estimation algorithms. \textit{Thirdly}, we design and conduct several real-world experiments involving uniform and nonuniform motions to evaluate the formation performance of our system. Experimental results show that our system performs considerably better than a benchmark system that uses EKF to estimate the leader’s states.

\section{Design of ROS-Compatible OISAC Scheme} \label{section: OISAC scheme design}
\begin{figure}
    \centering
    \includegraphics[width=3.4in]{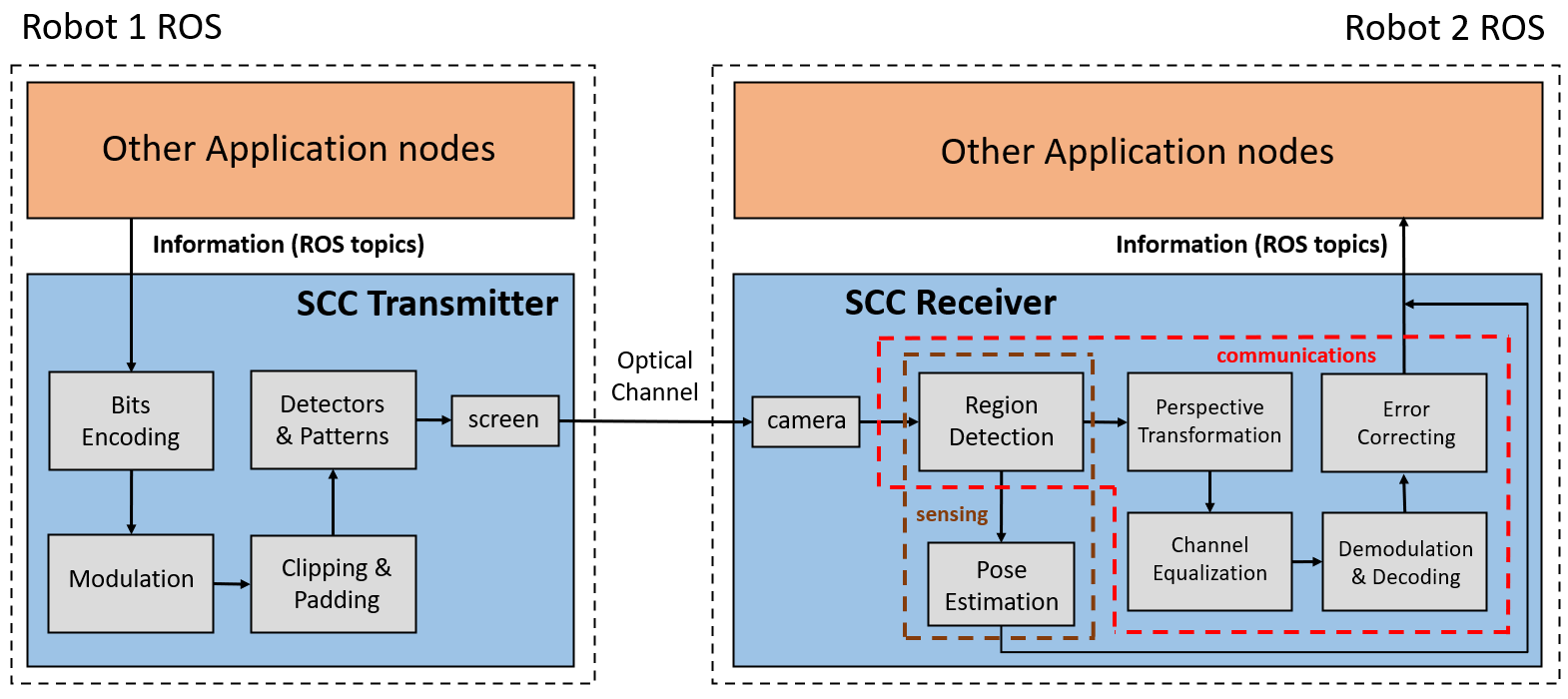}
    \caption{Block diagram of our ROS-compatible OISAC.}
    \label{fig: system overview}
    \vspace{-1.5em}
\end{figure}

In this section, we present our design of the ROS-compatible OISAC scheme. This scheme incorporates two parts: the camera sensing part, which estimates the relative pose between two robots, and the communication part implemented by SCC, which enables the leader to send its states to the follower. The block diagram of the proposed OISAC is shown in Fig.~\ref{fig: system overview}.
It is worth noting that existing SCC technologies designed for static scenarios may not function correctly on mobile robots with their default setups, as confirmed by our later experiments. Therefore, we will discuss the new problems posed by the considered mobile scenario and provide our solutions for overcoming them.
\subsection{Screen-Camera Link Design}
This subsection elaborates on the design and implementation of a screen-camera link to realize OISAC, which can be used to boost the cooperative performance of mobile robots. 

On the transmitter side, a commodity LCD screen is used to display visual information, and it continuously displays specific images, each containing a pixel matrix that carries data surrounded by four distinctive squares and white stripes, as illustrated in Fig.~\ref{fig: equipment setup}(b). Each square is considered a feature marker, and as we will explain later, the detection of these markers will be used for both camera sensing and pixel matrix extraction. Considering that objects with similar graphic features to the markers may exist in the background environments, we add a special graphic pattern (i.e., the black and white stripes) between the feature markers to enhance the detection accuracy. As shown in Fig.~\ref{fig: system overview}, the transmitted information is modulated into a two-dimensional (2-D) pixel matrix through a similar encoding and 2-D inverse Fourier Transform (IFFT) procedure presented in \cite{perli2010pixnet}. Meanwhile, each data bit is expanded by being interpolated with duplicate ones for high reliability. In our design, the first 4 bits are duplicated 5 times each and the second 4 bits are duplicated 3 times each. 

At the receiver side, a low-cost camera is fixed onboard the follower to capture RGB images, and we extract the information conveyed by the RGB images to enhance the control performance. To that end, an edge detection algorithm is applied to recognize and locate four feature markers. Specifically, we first convert each received image to a binary one and then apply the Canny edge detection algorithm \cite{canny1986computational} to extract graphic edges in the image and choose the edges with multi-hierarchy structure as potential candidates of the feature markers. The four feature markers are detected by verifying whether the pixels between the centers of each two candidates satisfy the predefined graphic pattern, where the centers of the markers are regarded as the feature points. The pixel coordinates of the feature points in the image can be used for camera sensing, i.e., relative pose estimation, which will be covered in Sec.~\ref{sec: pose estimation}. Moreover, other information, such as the transmitter's velocity, can also be sensed or estimated using various techniques like EKF and nonlinear velocity estimation\cite{mariottini2007leader, moshtagh2009vision}. On the other hand, the detection of the four feature points also contributes to SCC, as shown in Fig.~\ref{fig: system overview}. In practice, perspective distortion always exists in the received images since the screen and the camera can not be perfectly aligned. With the pixel coordinates of feature points, we conduct perspective transformation to correct the distortion so that the pixel matrix can be located and restored. To extract the information embedded in the pixel matrix, we apply demodulation and decoding operations in reverse to the aforementioned encoding and modulation process.

We conducted a series of experiments to evaluate the effects of distance and view angle on packet loss rates of the implemented SCC, where a low-cost Kinect camera and a 15.6-inch ThinkVision M14d monitor were used. The results are given in Table.~\ref{tab: PLR_distance} and \ref{tab: PLR_angle}, which show that our screen-camera link design has a packet loss rate of less than $3\%$ within 1.2-meter distance and 50-degree view angle.
We note that trade-offs exist between screen size and OISAC performance, as well as between sensing performance and communication performance. First, the larger screen size, the easier for the receiver to sense the feature points and decode the information bits. As such, the screen-camera link is able to achieve lower packet loss rate at longer distance and larger view angle. Nevertheless, a larger screen requires higher power consumption. Moreover, when considering a fixed screen size, a larger sensing area consisting of four feature markers and white stripes can enhance the sensing capability and accuracy. However, this also results in compressed data area (pixel matrix), which can lead to lower throughput and higher packet loss rates. To strike an optimal balance, it is essential to consider the specific requirements of the application at hand.
\vspace{-0.5em}
\begin{table}
    \begin{center}
        \caption{Packet loss rate under different distances.}
        \begin{tabular}{ccccccc}
            \toprule
            Distance (cm) & 50 & 60 & 70 & 80 & 90 & 100  \\
            \midrule
            Packet Loss Rate & 0.005 & 0.006 & 0.007 & 0.008 &	0.010 & 0.011 \\
            \bottomrule
            Distance (cm) & 110 & 120 & 130 & 140 & 150 \\
            \midrule
            Packet Loss Rate  & 0.017 & 0.028 & 0.057 & 0.102 & 0.357 \\
            \bottomrule
        \end{tabular}
        \label{tab: PLR_distance}
            \vspace{-1.1em}
    \end{center}
\end{table}
\begin{table}
    \begin{center}
        \caption{Packet loss rate under different view angles.}
        \begin{tabular}{cccccccc}
            \toprule
            View Angle (degree) & 0 & 10 & 20 & 30 & 40 & 50   \\
            \midrule
            Packet Loss Rate & 0.7\% & 0.8\% & 1\% & 1\% & 1.1\% & 1.3\% \\
            \bottomrule
        \end{tabular}
        \label{tab: PLR_angle}
    \end{center}
    \vspace{-1.1em}
\end{table}

\subsection{Problems in Implementing OISAC for Mobile Robots}
The application of OISAC on mobile robots was not straightforward, and we encountered two main problems in our experiments: 1) image blur caused by robot shaking, and 2) long image display delays. In this subsection, we will describe the problems we encountered and present our solutions to enable the application of OISAC on mobile robots.

Mobile robots often experience shaking due to mechanical constraints, especially when they start or change their velocity. This shaking can cause severe blurring of the images captured by onboard cameras, ultimately leading to failed sensing and communication. For instance, in our study case of a leader-follower formation control problem, the follower robot may fail to start moving at the beginning due to body shaking, which causes the captured images to be severely blurred, making it impossible for the follower to sense or receive any information, thus stopping immediately. In such cases, the follower may get caught in a starting-stopping cycle and shake violently in place, causing the formation to fail.
To mitigate this problem, we have implemented a velocity smoothing process to prevent drastic changes in the robots' velocity. Taking inspiration from the $keyboard\_teleop$ node in ROS, our velocity smoothing process involves constraining the robots' acceleration. Specifically, we denote $\mathbf{u}^{t} = [v^{t}, \omega^{t}]^T$ as the target velocity produced by the underlying control law, and $\mathbf{u}^{c} = [v^{c}, \omega^{c}]^T$ as the actual velocity of the robots. We then apply the following constraint:
\begin{equation}
\begin{aligned}
    \mathbf{u}^{c}_{i}(t) &= \begin{cases} \min \{ \mathbf{u}^{t}_{i}(t), \mathbf{u}^{c}_{i}(t - \delta_t) + \mathbf{a}_i\delta_t\}, \mathbf{u}^{t}_{i}(t) >  \mathbf{u}^{c}_{i}(t - \delta_t), \\ \max \{ \mathbf{u}^{t}_{i}(t), \mathbf{u}^{c}_{i}(t - \delta_t) - \mathbf{a}_i\delta_t\}, \mathbf{u}^{t}_{i}(t) \leq \mathbf{u}^{c}_{i}(t - \delta_t), \end{cases} \\
    i &= 1, 2.
\end{aligned}
\end{equation}
Here, $f_v = 1/\delta_t$ denotes the publishing frequency of the velocity topic in ROS, and $\mathbf{a} = [\dot{v}_{des}, \dot{\omega}_{des}]^T$ represents the absolute value of the desired acceleration. This process ensures steady movements, which significantly improves the quality of the received images. Additionally, it offers the advantage that if the receiver experiences bit errors and obtains incorrect information, the actual control signals will not change sharply, ensuring the overall system's robustness.

Apart from image blur, image display delay at the transmitter is another problematic issue. For example, in our experiments we found that the information displayed on the screen may be outdated. In existing throughput-oriented SCC designs, image display delay is not a major concern as they only focus on the number of packets received within a given duration and do not consider how stale the received packets are. However, in our screen-camera link, delay-sensitivity is crucial as real-time control is critical in collaborative mobile robotics applications. In ROS, messages are exchanged in the form of topics, and communications are achieved through topic publishing and subscribing under ROS protocols.
In our case, ROS nodes at the transmitter side publish topics on leader's states at a specific frequency, $f_{pub}$. The SCC transmitter node subscribes to the topic with a custom-sized queue of size $N_q$ and transforms the topics in the queue into displayed images. Assuming the time interval between the moment when the transmitter node receives its subscribed topic and when it displays the image embedding the topic information is $T_{tx}$, which mainly depends on the CPU's power and the size of the displayed image. If $f_{pub} > 1/T_{tx}$, indicating that the publishing frequency is higher than the image update frequency, the subscribing queue will be fully stacked, and the displayed images will be stale. We tested image display delays under different subscribing queue sizes, where the transmitter was running navigation with $f_{pub} = 20$~Hz, $T_{tx} = 60$~ms and displaying its current velocity information. As shown in Table~\ref{tab: delay under different queue sizes}, the display delay substantially increases as the subscribing queue size increases, which can significantly impact system performance. To achieve real-time communication and control, we set $N_q = 1$ by considering the fact that the follower is more concerned about the leader's latest states. In this sense, the image waiting to be displayed will be replaced by a newly generated image that incorporates fresher states of the leader.
\begin{table}
    \begin{center}
        \caption{Display delays under different queue sizes.}
        \begin{tabular}{ccccccc}
            \toprule
            Queue Size & 1 & 10 & 20 & 30 & 40 & 50  \\
            \midrule
            Average Delay (s) & 0.06 & 0.67 & 1.38 & 2.03 & 2.55 & 3.21 \\
            \bottomrule
        \end{tabular}
        \label{tab: delay under different queue sizes}
    \end{center}
\vspace{-1.2em}    
\end{table}
\section{Case Study: Vision-Based Leader-Follower Formation Control} \label{section: study case}
This section covers the leader-follower kinematics, camera model, visibility constraints, and problem formulation before detailing the relative pose estimation via camera sensing and the proposed control law. 
\subsection{Leader-Follower Kinematics}
\begin{figure} 
	\centering
	\subfloat[\label{fig: leader follower setup}]{
		\includegraphics[scale=0.17]{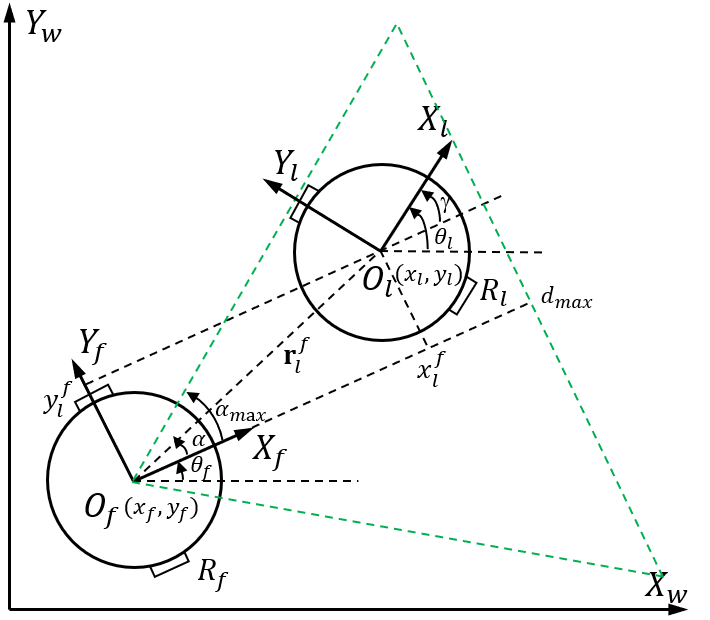}}
	\subfloat[\label{fig: camera model}]{
		\includegraphics[scale=0.17]{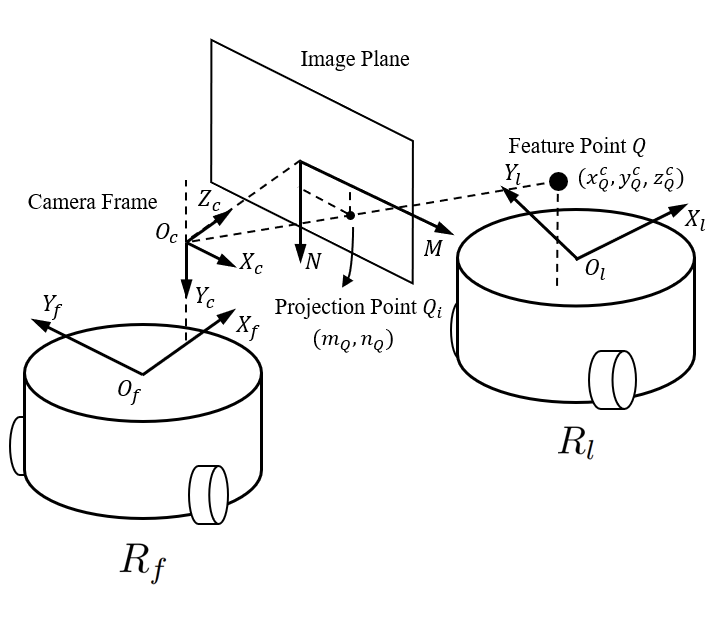}}
	\caption{(a) Leader-follower setup. (b) Camera model.}
	\label{fig1} 
	\vspace{-1.5em}
\end{figure}

As depicted in Fig. \ref{fig1}(a), we consider a vision-based leader-follower system consisting of two nonholonomic mobile robots $R_l$ and $R_f$, termed the leader and the follower, respectively. The follower is controlled to maintain a predefined relative pose to the leader based on the observations of the latter in its camera. The kinematics of each robot with respect to the world frame $\mathcal{W}$ can be written~as 
\begin{equation}
    \begin{bmatrix} \dot{x}_i \\ \dot{y}_i \\ \dot{\theta_i} \end{bmatrix} = \begin{bmatrix} \cos{\theta_i} & 0 \\ \sin{\theta_i} & 0 \\ 0 & 1 \end{bmatrix} \begin{bmatrix} v_i \\ \omega_i \end{bmatrix},
\end{equation} 
where $i \in \{ l, f \}$ refers to the leader or the follower, $\mathbf{r}_i = [x_i, y_i]^T$ and $\theta_i$ characterize the position and the orientation of robot $R_i$ in the world frame $\mathcal{W}$, and  $\mathbf{u}_i = [v_i, \omega_i]^T$ is the control input of robot $R_i$, including the linear velocity $v_i$ and the angular velocity $\omega_i$. 

Define the position of the leader $R_l$ with respect to the follower frame $\mathcal{F}$ as $\mathbf{r}_l^f = [x_l^f, y_l^f]^T$ to mathematically describe their relative position. We then have 
\begin{equation}
    \mathbf{r}_l^f = \begin{bmatrix} \cos{\theta_f} & \sin{\theta_f} \\ -\sin{\theta_f} & \cos{\theta_f} \end{bmatrix} (\mathbf{r}_l - \mathbf{r}_f).
\end{equation}

The time derivative of $\mathbf{r}_l^f$ is
\vspace{-0.5em}
\begin{equation}
    \dot{\mathbf{r}}_l^f = \begin{bmatrix} \dot{x}_l^f  \\ \dot{y}_l^f \end{bmatrix} = \begin{bmatrix} v_l\cos{\gamma} \\ v_l\sin{\gamma} \end{bmatrix}  + \begin{bmatrix} -1 & y_l^f \\ 0 & -x_l^f \end{bmatrix} \begin{bmatrix} v_f \\ \omega_f \end{bmatrix}, \label{eq:time_derivative_of_rlf}
\end{equation} 
where $\gamma = \theta_l - \theta_f$ is the relative orientation between $R_l$ and $R_f$ satisfying $\dot{\gamma} = \omega_l - \omega_f 
$. According to (\ref{eq:time_derivative_of_rlf}) and the definitions of $\gamma$ and $\dot{\gamma}$, the leader-follower kinematics can be expressed~as:
\begin{equation} \label{eq:kinematics}
\begin{aligned}
     \dot{\mathbf{s}} = \begin{bmatrix} \dot{x}_l^f  \\ \dot{y}_l^f \\ \dot{\gamma} \end{bmatrix} &= \begin{bmatrix} \cos{\gamma} & 0 \\ \sin{\gamma} & 0 \\ 0 & 1 \end{bmatrix} \mathbf{u}_l + \begin{bmatrix} -1 & y_l^f \\ 0 & -x_l^f \\ 0 & -1 \end{bmatrix} \mathbf{u}_f,
     \\ &= \mathbf{F}\mathbf{u}_l + \mathbf{G}\mathbf{u}_f,
\end{aligned}
\end{equation}  
where $\mathbf{s} = [x_l^f, y_l^f, \gamma]^T$ represents the relative pose between $R_l$ and $R_f$, and $\mathbf{u}_l = [v_l, \omega_l]^T$ and $\mathbf{u}_f = [v_f, \omega_f]^T$ are the control inputs of $R_l$ and $R_f$, respectively.

\subsection{Problem Statement} \label{subsection: problem statement}
Given a desired relative pose $\overline{\mathbf{s}} = [\overline{x}_l^f, \overline{y}_l^f, \overline{\gamma}]^T$, the objective is to design a controller for the follower so that the relative pose $\mathbf{s} = [x_l^f, y_l^f, \gamma]^T$ converges to an arbitrarily small neighborhood of the predefined $\overline{\mathbf{s}}$. Furthermore, the following assumptions are made in this paper for practical purposes.

\textit{Assumption 1:} The velocity and acceleration of the leader are bounded due to both mechanical constraints and visibility constraints, i.e., 
\begin{equation}
    |v_l| \leq v_{\max}, \ |\omega_l| \leq \omega_{\max}, \ |\dot{v}_l| \leq \dot{v}_{\max}, \ |\dot{\omega}_l| \leq \dot{\omega}_{\max} \label{eq: velocity constraints}.
\end{equation}

\textit{Assumption 2:} The leader is visible to the follower at the initial stage, and all feature points onboard the leader are detectable initially.

\textit{Assumption 3:} The relative orientation $\gamma$ is bounded to ensure that all feature points can be identified in the presence of perspective distortion, i.e.,
$    |\gamma| < \gamma_{\max} < \frac{\pi}{2}.$

\subsection{Camera Model and Visibility Maintenance} \label{subsection:camera model}
The camera mounted onboard the follower often has limited field of view (FoV). Assume that the optical axis of the camera is aligned with the forward direction of the follower, as shown in Fig.~\ref{fig1}(b). The pinhole camera model is used to project a 3-D point $\mathbf{r}_Q^c = [x_Q^c, y_Q^c, z_Q^c]^T$ in the camera frame to the image plane of the camera with coordinates $\mathbf{p}_Q^i = [m_Q, n_Q]^T$. The perspective projection is given by
\begin{equation}    
    m_Q = f_m x_Q^c/z_Q^c + m_0, \ \ n_Q = f_n y_Q^c/z_Q^c + n_0,
\end{equation} 
where $f_m$ and $f_n$ are pixel scaling factors; $(m_0, n_0)$ is the image coordinates of the camera's principal point. 

Since the camera has limited visual capability, the leader is visible to the follower only if the leader is within the follower's FoV. Assume that the visible region of the follower is a cone whose centerline coincides with the optical axis of the camera, as shown in the green dashed area in the Fig.~\ref{fig1}(a). Define $\alpha = \arctan(y_l^f/x_l^f)$ as the bearing angle of the leader's center with respect to the follower frame, and  $\alpha_{\max}$ and $d_{\max}$ as the maximum angle and distance of view, respectively. To maintain the visual observation of the leader in the FoV of the follower, the following conditions should be satisfied:
\begin{equation} 
\begin{aligned}
        2\mu\cos{(\alpha_{\max}-\frac{\pi}{6})} < x_l^f \leq d_{\max} - \mu, \\
        \ |\alpha| \leq \arctan\left( \frac{x_l^f\sin{\alpha_{\max}} - \mu}{x_l^f\cos{\alpha_{\max}}}\right),
\end{aligned}
\end{equation}
where $\mu$ is the collision radius of the leader. 

In our study case, we use the OISAC scheme to transmit the leader's velocity information to the follower. Define $\hat{\mathbf{u}}_l = [\hat{v}_l, \hat{\omega}_l]^T$ as the obtained velocity information from the SCC link, and $\bm{\delta} = [\delta_v, \delta_{\omega}]^T = [v_l - \hat{v}_l, \omega_l - \hat{\omega}_l]^T$ as the error between $\mathbf{u}_l$ and $\hat{\mathbf{u}}_l$. Normally $\bm{\delta}$ is bounded by the quantization error $[v_{max}/2^{n+1}, \omega_{max}/2^{n+1}]^T$, where the number of bits is set to $n = 8$ in this paper. With the acceleration constraints described in (\ref{eq: velocity constraints}), $\hat{\mathbf{u}}_l$ should not change drastically in a short time interval $\triangle t$. We thus have 
\begin{align}
    |\hat{v}_l(t) - \hat{v}_l(t - \triangle t)| & \leq N\dot{v}_{des}\triangle t \label{eq: acc_v constraints}, \\
    |\hat{\omega}_l(t) - \hat{\omega}_l(t - \triangle t)| & \leq N\dot{\omega}_{des}\triangle t
    \label{eq: acc_w constraints},
\end{align}
where $N$ is a positive integer. In case $\hat{\mathbf{u}}_l(t)$ does not satisfy the acceleration constraints (\ref{eq: acc_v constraints})-(\ref{eq: acc_w constraints}), or the feature points are not detected correctly, the corresponding visual information captured in the previous sampling interval will be applied. This verification process implies that $\bm{\delta}$ is bounded~by
\begin{align}
    \delta_v = |v_l - \hat{v}_l| &\leq \delta_v^+ = \max \left\{ \frac{v_{max}}{2^{n+1}}, N\dot{v}_{des}\triangle t \right\} \label{eq: v error bound}, \\
    \delta_\omega = |\omega_l - \hat{\omega}_l| &\leq \delta_\omega^+ = \max \left\{ \frac{\omega_{max}}{2^{n+1}}, N\dot{\omega}_{des}\triangle t \right\} \label{eq: w error bound}.
\end{align}
where the period interval $\triangle t$ is chosen to be $100$ms. Furthermore, substituting (\ref{eq: velocity constraints}) into (\ref{eq: v error bound})-(\ref{eq: w error bound}) yields
\begin{align}
    |\hat{v}_l| &\leq \hat{v}_l^+ = v_{max} + N\dot{v}_{des}\triangle t, \\
    |\hat{\omega}_l| &\leq \hat{\omega}_l^+ = \omega_{max} + N\dot{\omega}_{des}\triangle t \label{eq: obtained w bound}.
\end{align}


\subsection{Camera Sensing: Relative Pose Estimation} \label{sec: pose estimation}
We now describe our camera sensing method for the follower to estimate the relative pose from the leader using only the RGB images captured by the onboard camera. In most vision-based leader-follower formation control methods, typically the leader is equipped with particular markers so that the follower can position them in the camera. Here our markers are the aforementioned four feature points displayed on the screen.

\begin{figure}
    \centering
    \includegraphics[width=2.8in]{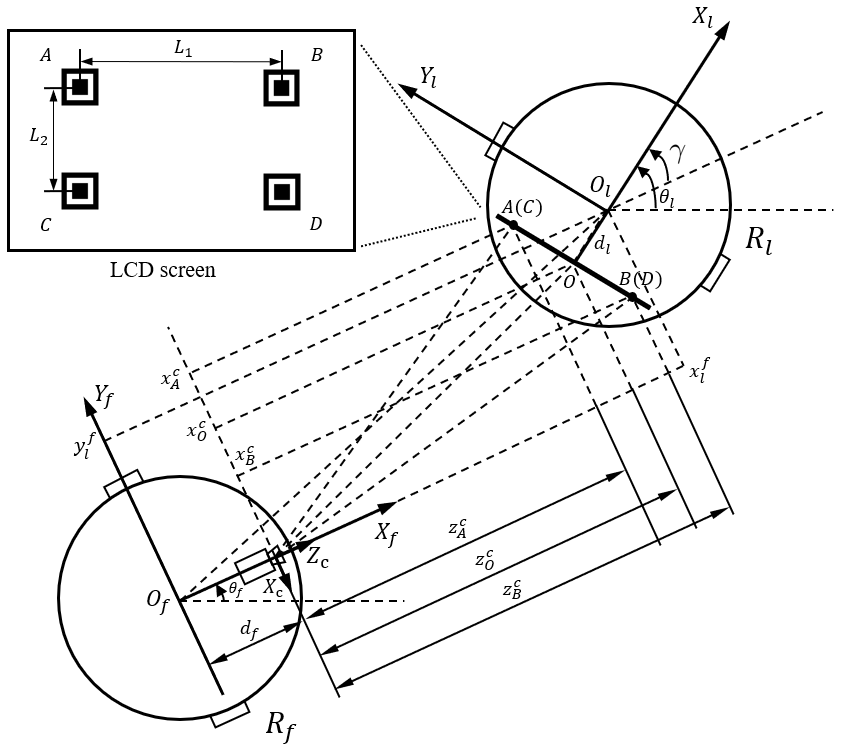}
    \caption{Pose estimation.}
    \label{pose estimation}
    \vspace{-1.5em}
\end{figure}

Assume that the plane of the screen is perpendicular to the forward direction of the leader, and the four feature points form a rectangle. The camera onboard the follower satisfies the pinhole camera model discussed in Sec. \ref{subsection:camera model}. As shown in Fig.~\ref{pose estimation}, we assume that the horizontal distance between the origin of the follower frame and the camera's principal point is $d_f$, and the horizontal distance between the origin of the leader frame and the LCD screen is $d_l$. The feature points are horizontally separated by $L_1$ and vertically separated by $L_2$. Both robots have a prior knowledge of $d_f$, $d_l$, $L_1$ and $L_2$. From the observation of the camera, the pixel coordinates of the feature points $A$, $B$, $C$ and $D$ in the image plane are denoted by  $\mathbf{p}_A^i = [m_A, n_A]^T$, $\mathbf{p}_B^i = [m_B, n_B]^T$, $\mathbf{p}_C^i = [m_C, n_C]^T$, $\mathbf{p}_D^i = [m_D, n_D]^T$, respectively. Define $\widetilde{m}_j = m_j - m_0$, $\widetilde{n}_j = n_j - n_0$, $j = \{ A, B, C, D \}$ as the pixel distances between the $i$-th pixel point and the origin of the image plane. Based on the pixel coordinates that can be directly acquired from the received RGB image, we can reconstruct the following coordinates $(z_A^c,x_A^c)$ and $(z_B^c,x_B^c)$ of feature points $A$ and $B$ in the camera frame, where 
\vspace{-0.25em}
\begin{align}
    z_A^c &= \frac{f_n L_2}{n_C - n_A}, \  z_B^c = \frac{f_n L_2\widetilde{n_A}}{(n_C - n_A)\widetilde{n_B}}, \label{eq: pixel to camera} \\
    x_A^c &= \frac{f_n L_2\widetilde{m_A}}{f_m(n_C - n_A)}, \ x_B^c = \frac{f_n L_2\widetilde{m_B}\widetilde{n_A}}{f_m(n_C - n_A)\widetilde{n_B}}. 
\end{align}
\vspace{-0.75em}

According to the following geometric relationship depicted in Fig. \ref{pose estimation}, the coordinates of the screen's center $O$ in the camera frame can be derived as:
\vspace{-0.25em}
\begin{align}
    z_O^c &= \frac{1}{2}(z_A^c + z_B^c) = \frac{f_n L_2(\widetilde{n_A} + \widetilde{n_B})}{2(n_C - n_A)\widetilde{n_B}},  \\
    x_O^c &= \frac{1}{2}(x_A^c + x_B^c) = \frac{f_n L_2(\widetilde{m_A}\widetilde{n_B} + \widetilde{m_B}\widetilde{n_A})}{2f_m(n_C - n_A)\widetilde{n_B}}.
\end{align}
\vspace{-0.75em}
    
The relative distance $\mathbf{r}_l^f = [x_l^f, y_l^f]^T$ between the leader and the follower can then be estimated~as:
\vspace{-0.25em}
\begin{align}
    x_l^f &= z_O^c + d_f + d_l\cos{\gamma}, \\
    y_l^f &= -x_O^c + d_l\sin{\gamma},
\end{align}
where the relative orientation $\gamma$ satisfies
\begin{align}
    \sin{\gamma} &= \frac{1}{L_1}(z_B^c - z_A^c) = \frac{f_n L_2(\widetilde{n_A} - \widetilde{n_B})}{L_1(n_C - n_A)\widetilde{n_B}}, \\
    \cos{\gamma} &= \frac{1}{L_1}(x_B^c - x_A^c) = \frac{f_n L_2(\widetilde{m_B}\widetilde{n_A} - \widetilde{m_A}\widetilde{n_B})}{f_m L_1(n_C - n_A)\widetilde{n_B}} \label{eq: gamma formula}.
\end{align}

\subsection{Control Law} \label{sec: control law}
In this subsection, we devise a control law that can leverage the leader's velocity information and the relative pose extracted from the OISAC scheme to achieve a stable formation between the two robots.

Define the vector of formation errors as $\bm{\varepsilon} = [\varepsilon_x, \varepsilon_y, \varepsilon_{\gamma}]^T = [x_l^f - \overline{x}_l^f, y_l^f - \overline{y}_l^f, \gamma - \overline{\gamma}]^T$.
To maintain a desired formation is equivalent to ensuring that $\bm{\varepsilon}$ converges to an arbitrarily small neighborhood of $\mathbf{0} \in \mathbb{R}^3$. Recall that the time derivative of $\bm{\varepsilon}$ has been  given in (\ref{eq:kinematics}). Define $\sigma = 1/\left( \left(x_l^f \right)^2 + 1 \right)$. We propose the following control law:
\begin{equation} \label{eq: control law}
    \mathbf{u}_f = \begin{bmatrix} v_f \\ \omega_f \end{bmatrix} = \sigma\mathbf{H} (\mathbf{K}\bm{\varepsilon} + \mathbf{F}\hat{\mathbf{u}_l}),
\end{equation} 
where $\mathbf{K} = \mathrm{diag}(k_1, k_2, k_3)$ contains three tunable positive scalars; recall that $\hat{\mathbf{u}_l} = [\hat{v}_l, \hat{\omega}_l]^T$ is the velocity information captured by the follower using the vision scheme presented in Sec.~\ref{section: OISAC scheme design}; the matrix $\mathbf{H}$ is given by
\begin{equation} \label{eq: matrix H}
    \mathbf{H} = \begin{bmatrix} 1/\sigma & x_l^f y_l^f & y_l^f \\ 0 & x_l^f & 1 \end{bmatrix}.
\end{equation}

We have the following theorem regarding the formation errors of the control law in (\ref{eq: control law}):

\textit{\textbf{Theorem 1}}: Considering a leader-follower system with the kinematics in  (\ref{eq:kinematics}) satisfying Assumptions 1-4 and the formation control law in (\ref{eq: control law}), the prescribed stable formation performance in Sec.~\ref{subsection: problem statement} can be achieved by properly selecting parameters $k_1$, $k_2$ and $k_3$. That is, the formation errors $\bm{\varepsilon}$ are bounded and have guaranteed convergences. 

\textit{Proof:} Consider the following Lyapunov candidate $V = \bm{\varepsilon}^T\bm{\varepsilon}/2$. Denote $\eta = x_l^f \sigma$. Substituting (\ref{eq:kinematics}) and (\ref{eq: control law})-(\ref{eq: matrix H}) into the Lyapunov candidate, we have its time derivative given~by
\vspace{-0.3em}
\begin{equation}
\begin{aligned}
    \dot{V} =& \bm{\varepsilon}^T\dot{\bm{\varepsilon}} = \bm{\varepsilon}^T[\mathbf{F}\mathbf{u}_l + \sigma\mathbf{G}\mathbf{H} (\mathbf{K}\bm{\varepsilon} + \mathbf{F}\hat{\mathbf{u}_l})], \\
    =& -k_1\varepsilon_x^2 - k_2 \eta x_l^f \varepsilon_y^2 - k_3\sigma\varepsilon_{\gamma}^2 + \varepsilon_x(v_l - \hat{v}_l)\cos{\gamma}  \\ 
    &+ \varepsilon_y[(v_l - \eta x_l^f \hat{v}_l)\sin{\gamma} -\eta\hat{\omega}_l - \eta(k_2 + k_3)\varepsilon_{\gamma}]  \\
    &+ \varepsilon_{\gamma}(\omega_l - \sigma\hat{\omega}_l - \eta\hat{v}_l\sin{\gamma}) \label{eq: Lypuanove theory}.
\end{aligned}
\end{equation}
\vspace{-0.5em}

For clarity, we define
\vspace{-0.5em}
\begin{align}
    \xi_1 &= (v_l - \hat{v}_l)\cos{\gamma} \label{eq: p1}, \\
    \xi_2 &= (v_l - \eta x_l^f \hat{v}_l)\sin{\gamma} -\eta\hat{\omega}_l - \eta(k_2 + k_3)\varepsilon_{\gamma},\\
    \xi_3 &= \omega_l - \sigma\hat{\omega}_l - \eta\hat{v}_l\sin{\gamma} \label{eq: p3}.
\end{align}

\vspace{-0.25em}
Substituting (\ref{eq: v error bound})-(\ref{eq: obtained w bound}) into (\ref{eq: p1})-(\ref{eq: p3}) further yields
\vspace{-0.25em}
\begin{align}
    |\xi_1| &\leq \delta_v^+, \\
    |\xi_2| & \leq |v_l - \hat{v}_l| + |\sigma\hat{v}_l| + \eta[(k_2 + k_3)|\varepsilon_{\gamma}| + |\hat{\omega}_l|] \nonumber \\
    & \leq \delta_v^+ + \sigma\hat{v}_l^+ + \eta[(k_2 + k_3)|\varepsilon_y| + \hat{\omega}_l^+],  \\
    |\xi_3| & \leq |\omega_l - \hat{\omega}_l| + |\eta x_l^f\hat{\omega}_l| + |\eta\hat{v}_l| \nonumber\\
    & \leq \delta_\omega^+ + \eta x_l^f\hat{\omega}_l^+ + \eta\hat{v}_l^+  .
\end{align}

By properly choosing $k_1$, $k_2$ and $k_3$ such that $k_1 \geq 2\delta_v^+/|\varepsilon_x|$, $k_2 \geq 2[\delta_v^+ + \sigma\hat{v}_l^+ + \eta(k_3|\varepsilon_y| + \hat{\omega}_l^+)]/(\eta (x_l^f+1) |\varepsilon_y|)$, and $k_3 \geq 2(\delta_\omega^+ + \eta x_l^f\hat{\omega}_l^+ + \eta\hat{v}_l^+)/(\sigma|\varepsilon_\gamma|)$, we can bound $\dot{V}$ given in  (\ref{eq: Lypuanove theory}) as follows
\begin{equation}
\begin{aligned}
    \dot{V} &\leq -(k_1\varepsilon_x^2 + k_2 \eta x_l^f \varepsilon_y^2 + k_3\sigma\varepsilon_{\gamma}^2)/2 \\
    &\leq -\phi\bm{\varepsilon}^T\bm{\varepsilon}/2 = -\phi V  \label{eq: Lyapunov proof},\\
\end{aligned}
\end{equation}
where $\phi = \min\{ k_1, k_2\eta x_l^f, k_3\sigma \}$. According to the Lyapunov stability theory, (\ref{eq: Lyapunov proof}) indicates that the proposed leader-follower system is asymptotically stable, and the error $\bm{\varepsilon}$ is bounded. This ends the proof. 

\section{Experimental Results} \label{section: experiments}
\begin{figure*} 
	\centering
	   \subfloat[\label{fig: circle tra}]{
		 \includegraphics[scale=0.35]{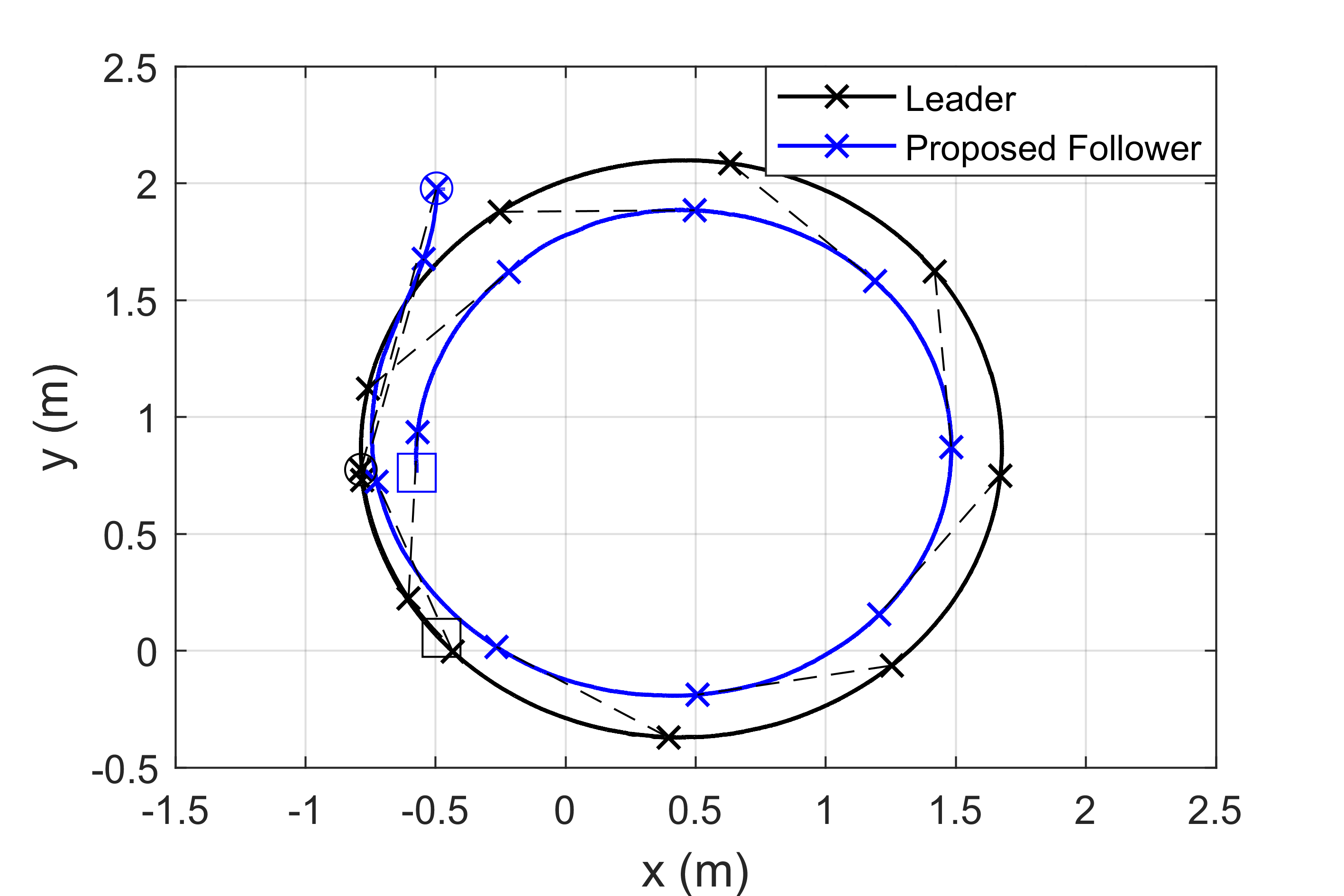}}
	   \subfloat[\label{fig: circle ex}]{
		 \includegraphics[scale=0.35]{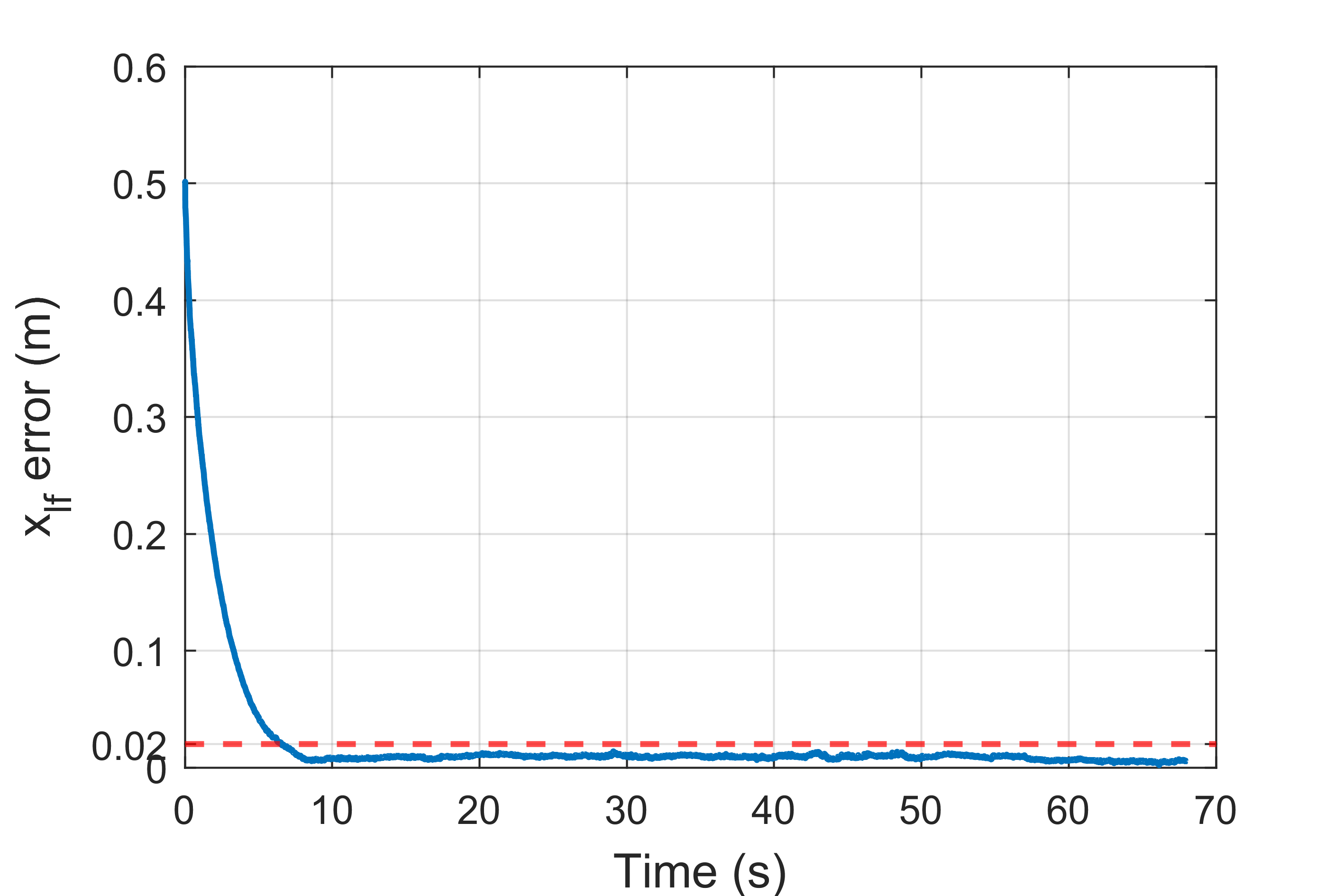}}
          \subfloat[\label{fig: circle ey}]{
		 \includegraphics[scale=0.35]{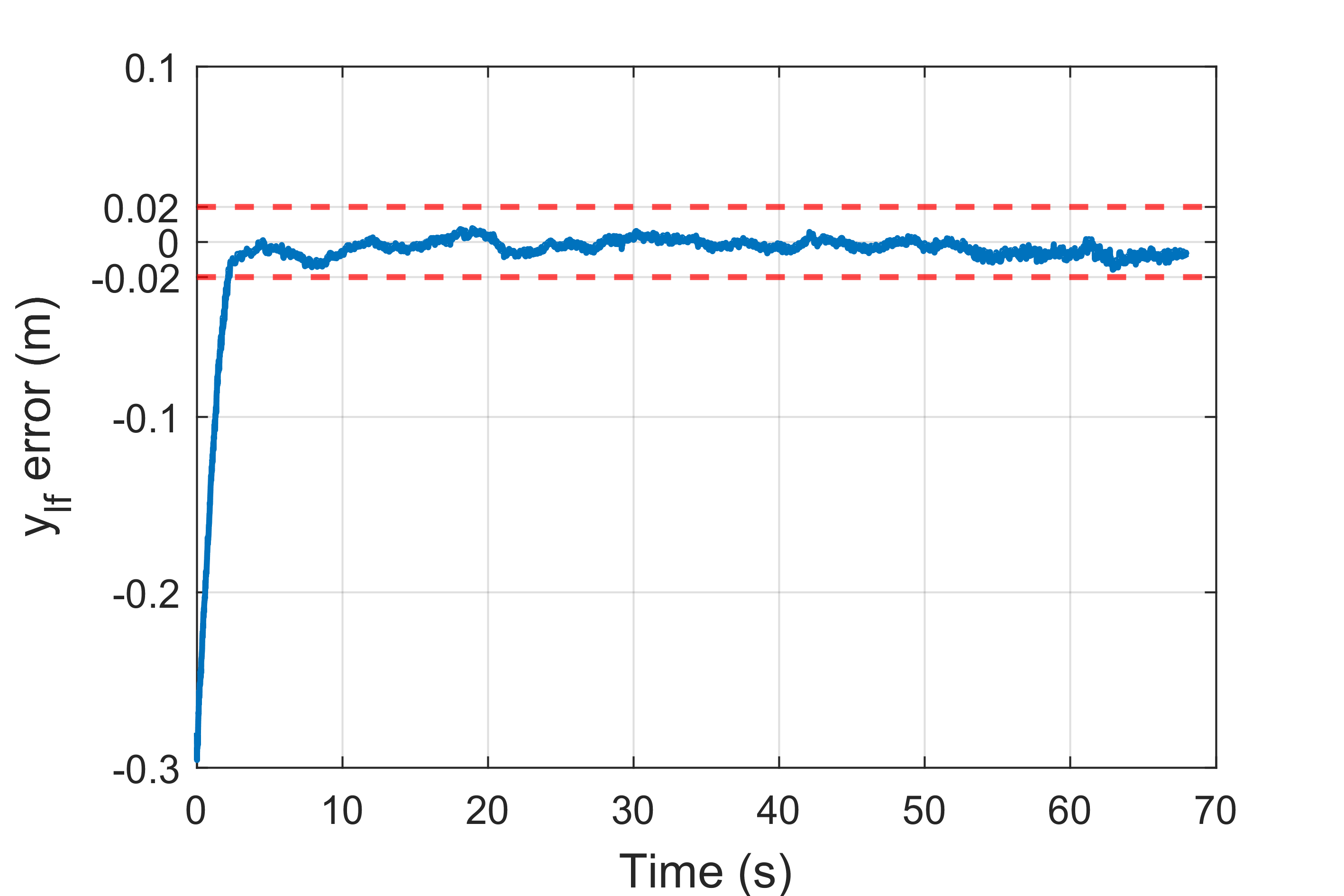}}
          \subfloat[\label{fig: circle et}]{
		 \includegraphics[scale=0.35]{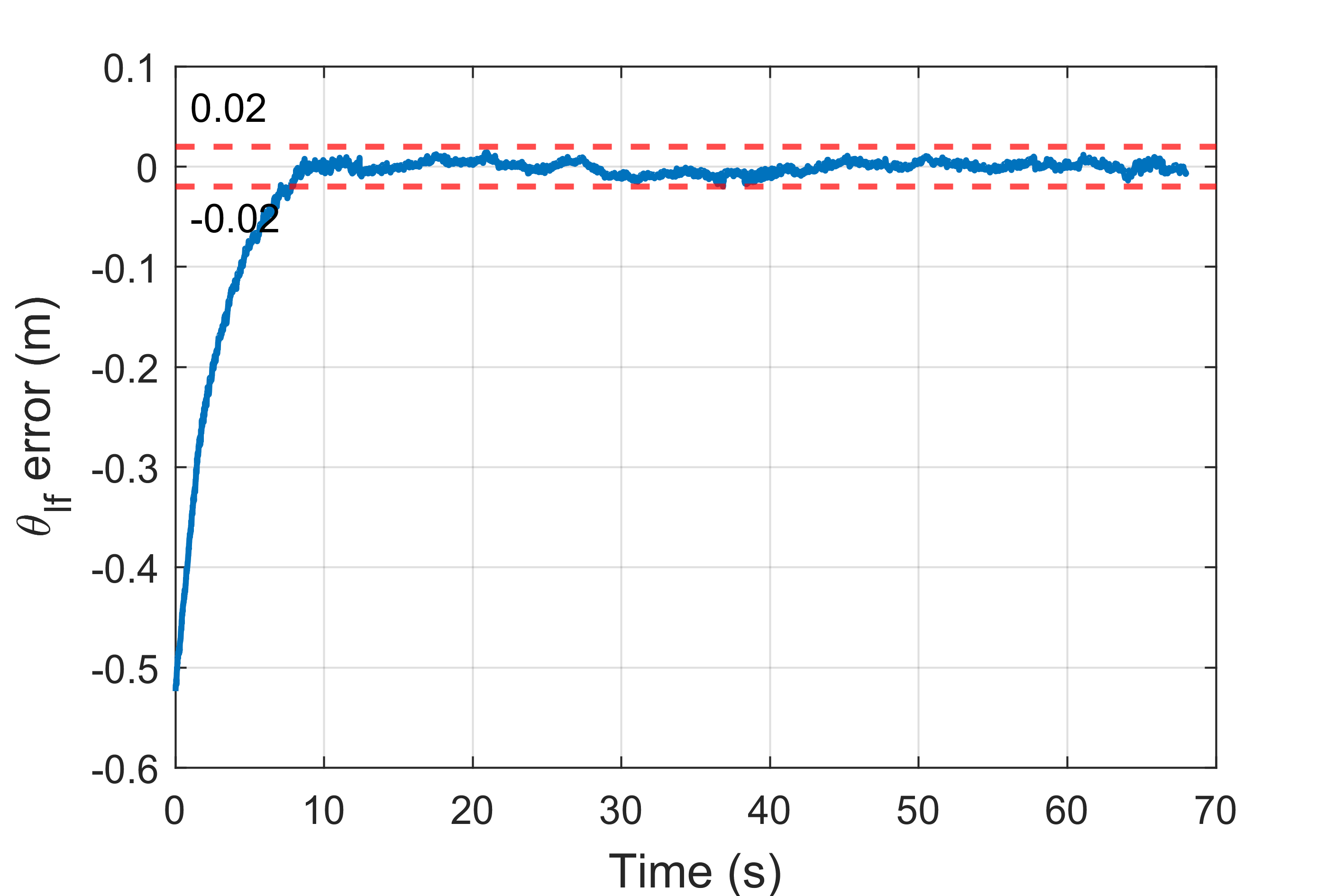}}
	\caption{Results of the first experiment. (a) Trajectory. (b)-(d) Formation errors.}
	\label{fig: results of the circle path} 
	\vspace{-1.5em}
\end{figure*}
We conducted real-world experiments on two Turtlebot2 robots to evaluate the proposed formation system. 
In our experiments, the VICON motion capture system is used to obtain the positions of the mobile robots with respect to a global frame that are used as ground truth. The frequency of the VICON cameras is set to 100 frames per second (fps). We attached three reflective markers on the top layer of each robot forming an isosceles triangle so that the orientation of each robot can be determined by the coordinates of three markers. The follower is equipped with a Kinect camera to capture RGB images. Each robot is connected to an Intel NUC mini PC, running on Ubuntu 18.04. The vision algorithm and the proposed control law are implemented in robot operating system (ROS) Melodic. For robustness and practical purposes, the parameter settings are chosen as follows: $d_{\max} = 1.45$m, $\gamma_{\max} = \pi/3$, $\alpha_{\max} = \pi/4$, $r = 0.2$m, $v_{\max} = 0.6$m/s, $\omega_{\max} = 0.2$rad/s, $\dot{v}_{\max} = 0.5 \rm{m/s^2}$, $\dot{\omega}_{\max} = 0.2 \rm{rad/s^2}$, $k_1 = 0.5$, $k_2 = 0.75$, $k_3 = 0.5$, $N$ = 5, $\triangle t = 100$ms, $f_m = f_n = 500$pixels/m, $m_0 = 320$pixels, $n_0 = 240$pixels, $L_1 = 0.232$m, $L_2 = 0.145$m, $d_l = 0.275$m, $d_f = -0.017$m. To evaluate the formation stability and robustness, we conducted the following three experiments: 1) formation along a straight/circular path with the velocity of the leader being constant; 2) braking distance when the leader sharply decelerates; 3) formation along a U-shaped path with velocity of the leader being time-varying. 

The first experiment is designed to test the basic formation performance of the proposed scheme, where the leader's velocity is set to be constant. Due to space limitation, we only show the experimental results of the case with a circular path, though even better performance has been observed for the case with a straight path. 
In this experiment, the leader moves along a circular path with $v_l = 0.125$m/s and $\omega_l = 0.1$rad/s. The relative pose is initialized as $\mathbf{s}_0 = [1.25, -0.3, 0]^T$, and the desired one is set to $\overline{\mathbf{s}} = [0.75, 0, \pi/6]^T$. The experimental results are shown in Fig. \ref{fig: results of the circle path}. The trajectories recorded by the motion capture system are depicted in Fig. \ref{fig: results of the circle path}(a), where the circles mark the starting points and the squares mark the ending points. Figs. \ref{fig: results of the circle path}(b)-\ref{fig: results of the circle path}(d) show that the formation errors quickly reduce and then varies within a small range, i.e., $\hat{\bm{\varepsilon}} = [\pm 0.02, \pm 0.02, \pm 0.02]^T$. The results indicate that the proposed scheme can achieve a stable and accurate formation under linear/circular motion with a constant velocity. 


\begin{figure}
    \centering
    \includegraphics[width=3in]{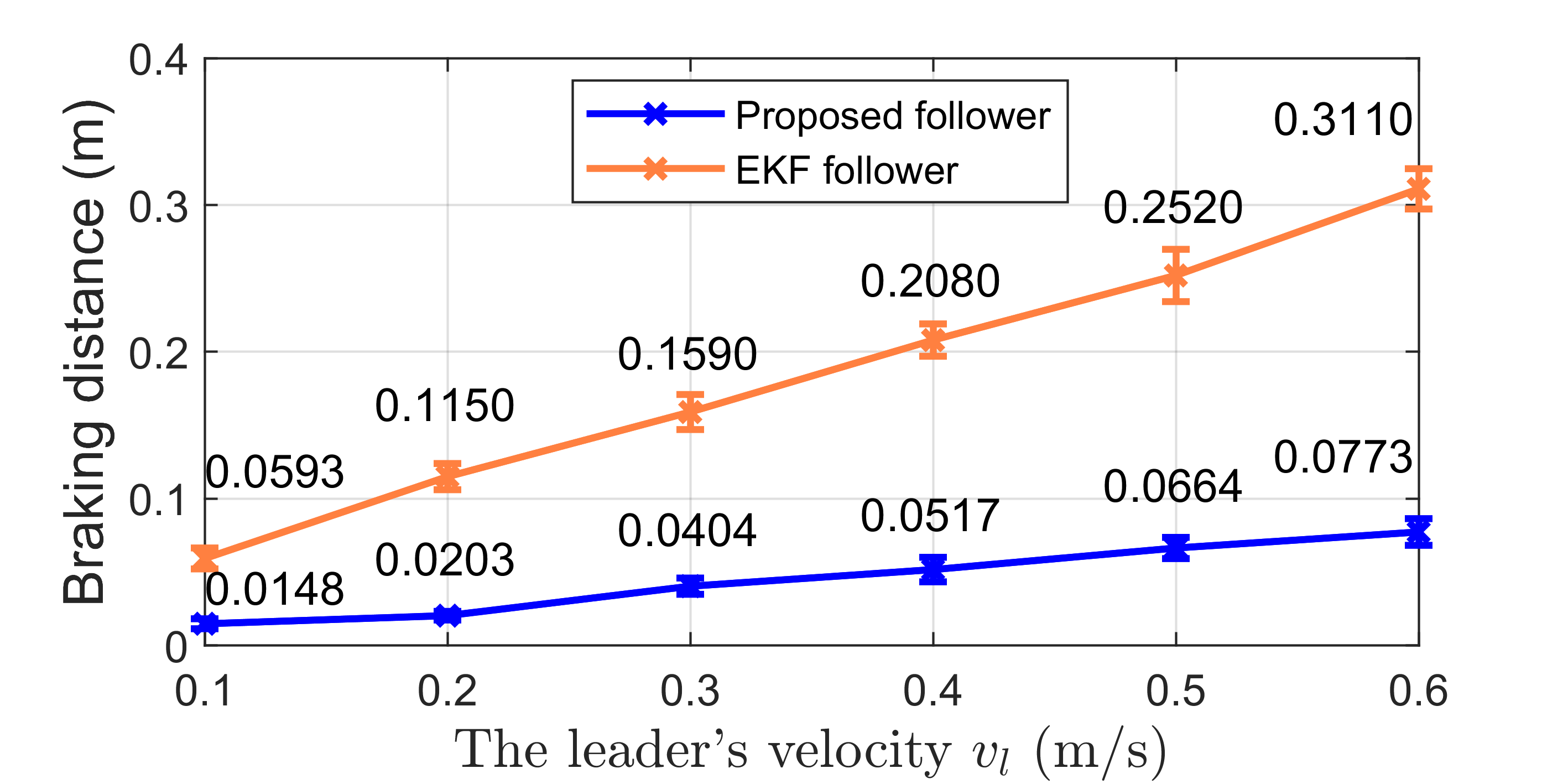}
    \caption{Braking distance in the second experiment.}
    \label{fig: braking distance}
    \vspace{-1.5em}
\end{figure}  
In order to demonstrate the merits of our proposed system, we designed experiments involving motions with dynamic velocity of the leader (i.e., $v_l$ may change). Specifically, in the second experiment, the leader executed a linear motion until the formation stabilizes, and then sharply braked with the maximum deceleration. This experiment can evaluate the follower's responsiveness to the leader's dynamic velocity. Thanks to the OISAC scheme developed in Sec.~\ref{section: OISAC scheme design}, the follower is expected to react faster to the leader's deceleration to avoid significant formation errors or even a collision. We measured the braking distance of the follower at six levels of $v_l$ from 0.1m/s to 0.6m/s. Additionally, we implemented the velocity estimation method using extended Kalman filtering (EKF) \cite{das2002vision} to serve as the benchmark. For each level of $v_l$, 10 measurements are performed for both the proposed method and the benchmarking EKF method, totaling 120 measurements. We averaged the 10 measurement results for each $v_l$ as the performance metric. The results are presented in Fig. \ref{fig: braking distance}. It can be clearly observed that compared with the follower using the EKF method, the follower based on the proposed vision scheme has much shorter braking distance (all not exceeding $0.08$m), about 3$\times$ to 5$\times$ shorter than that of the benchmark. This is because our OISAC-augmented follower is more agile to the drastic changes of the leader's velocity, resulting in a more robust control law in nonuniform motions. 

\begin{figure} 
	\centering
	   \subfloat[\label{fig: u tra}]{
		 \includegraphics[scale=0.35]{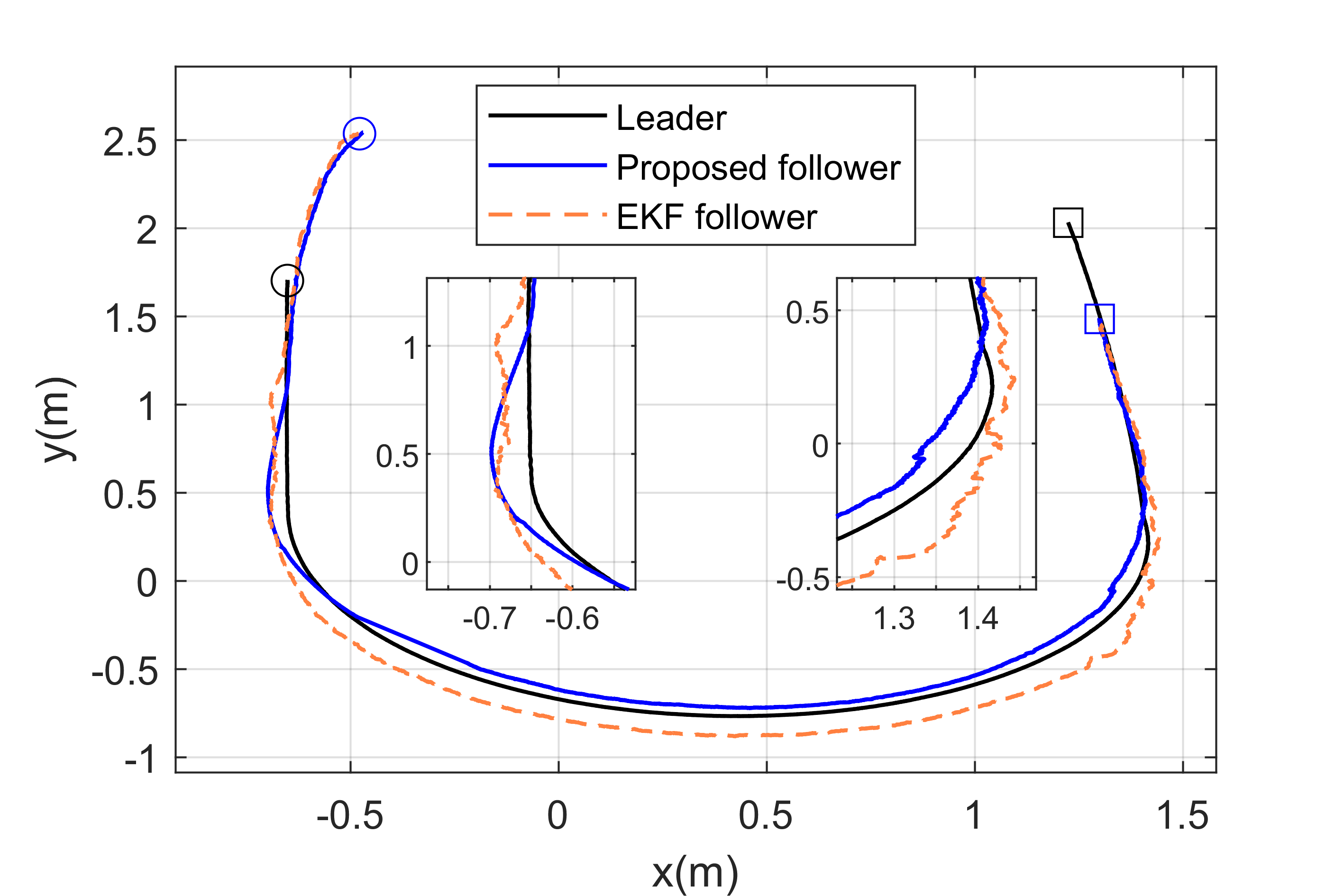}}
   \vspace{-1em}
	   \subfloat[\label{fig: u ex}]{
		 \includegraphics[scale=0.35]{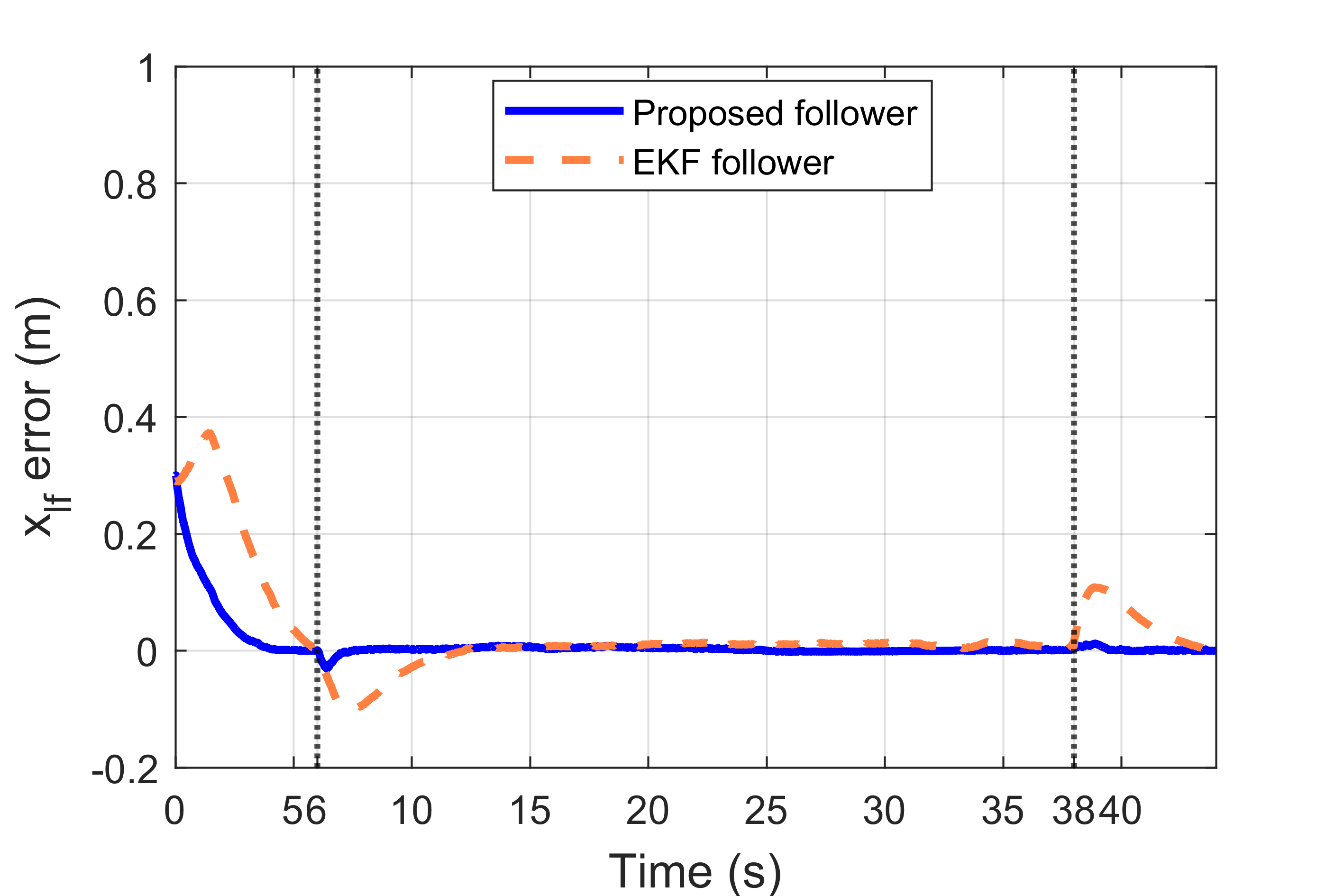}}
        \\
          \subfloat[\label{fig: u ey}]{
		 \includegraphics[scale=0.35]{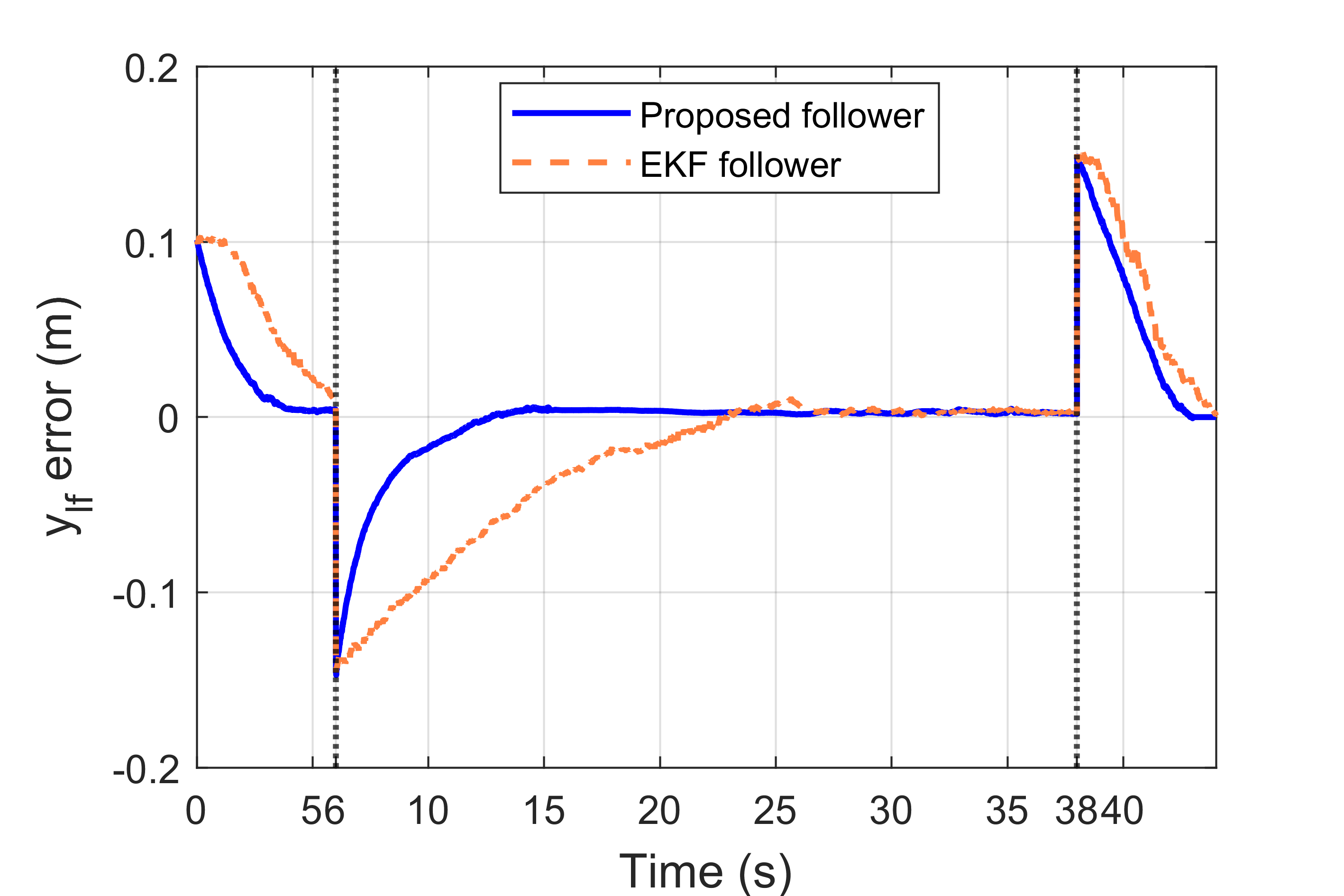}}
   \vspace{-1em}
          \subfloat[\label{fig: u et}]{
		 \includegraphics[scale=0.35]{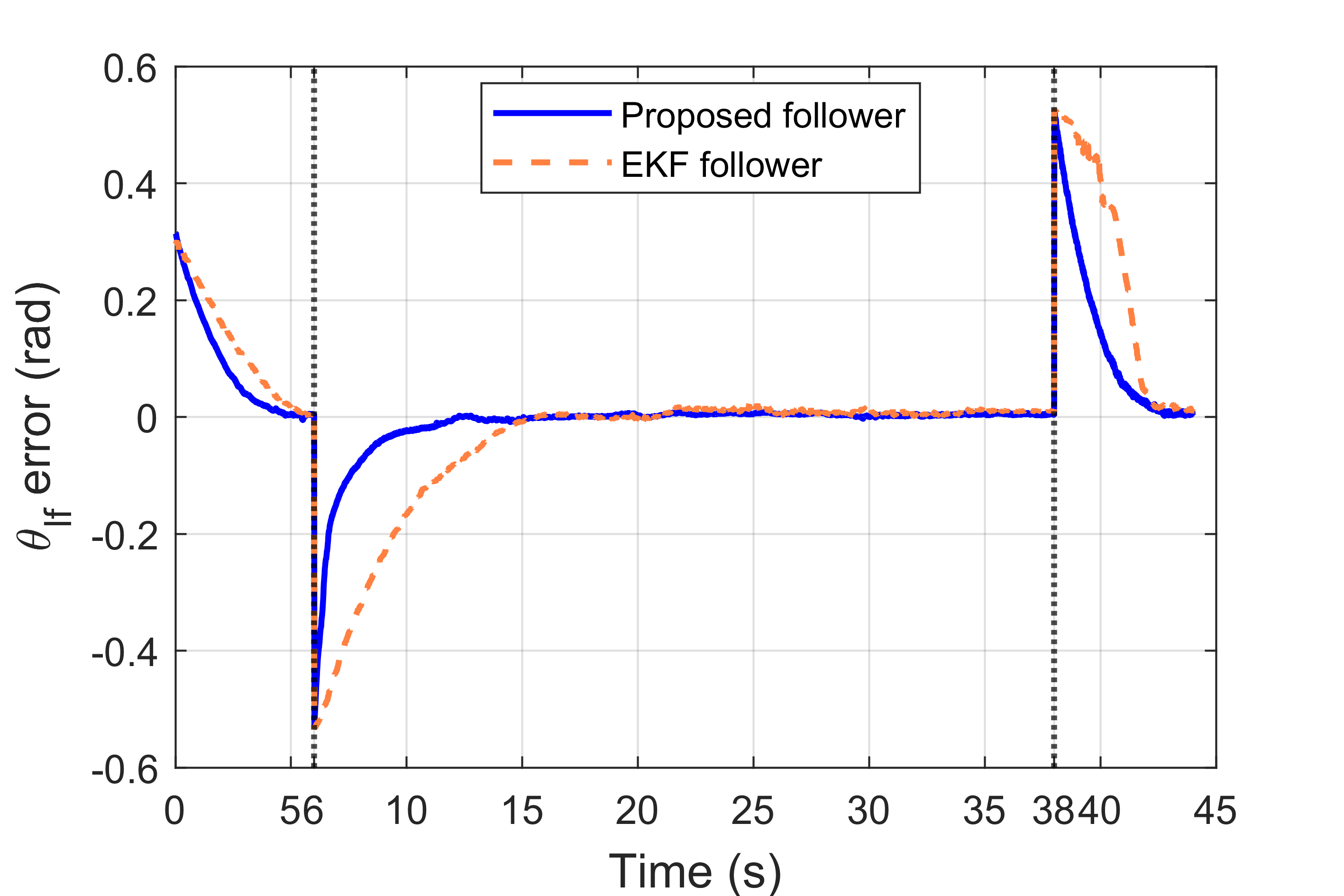}}
        \\
          \subfloat[\label{fig: u linear}]{
		 \includegraphics[scale=0.35]{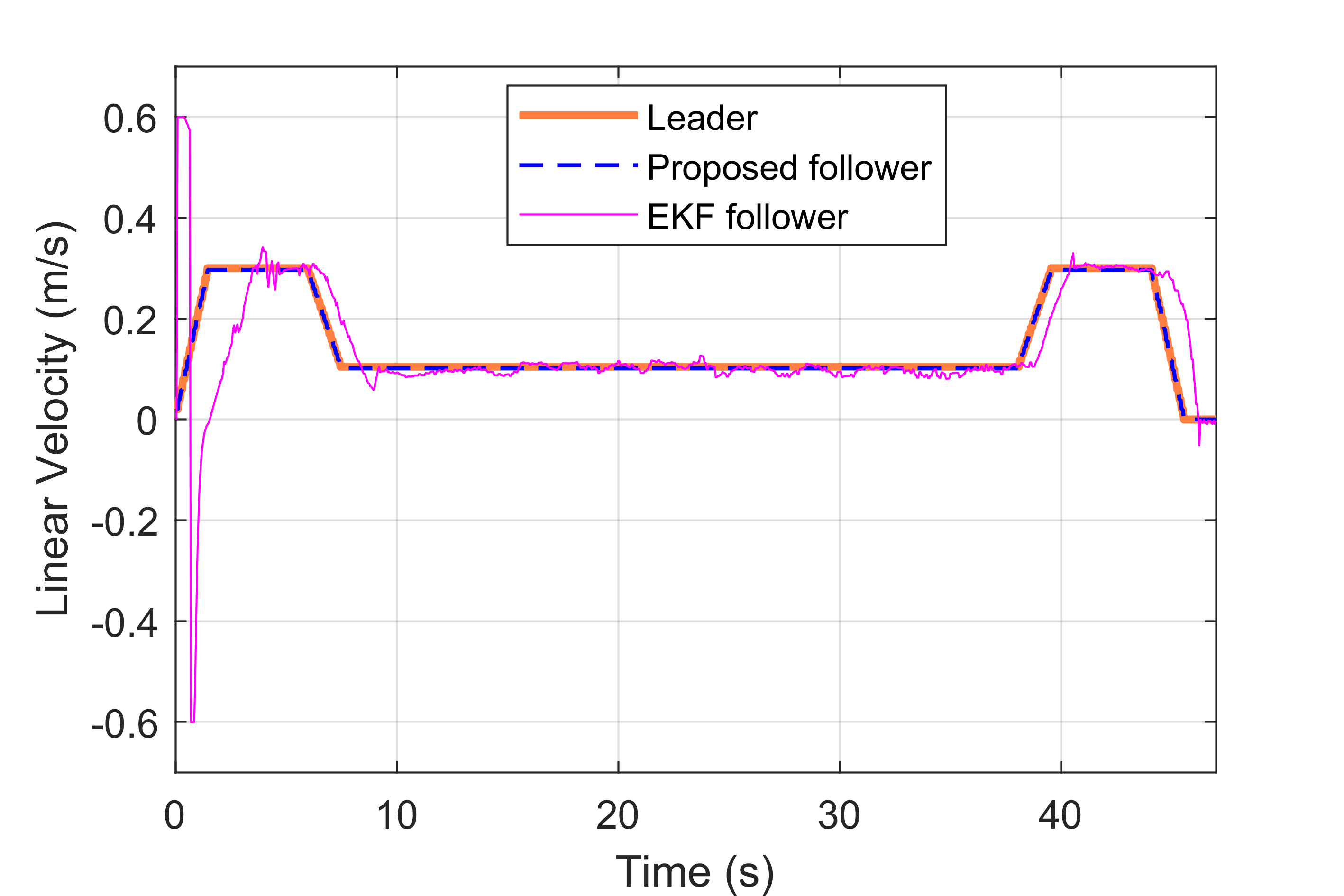}}
          \subfloat[\label{fig: u angular}]{
		 \includegraphics[scale=0.35]{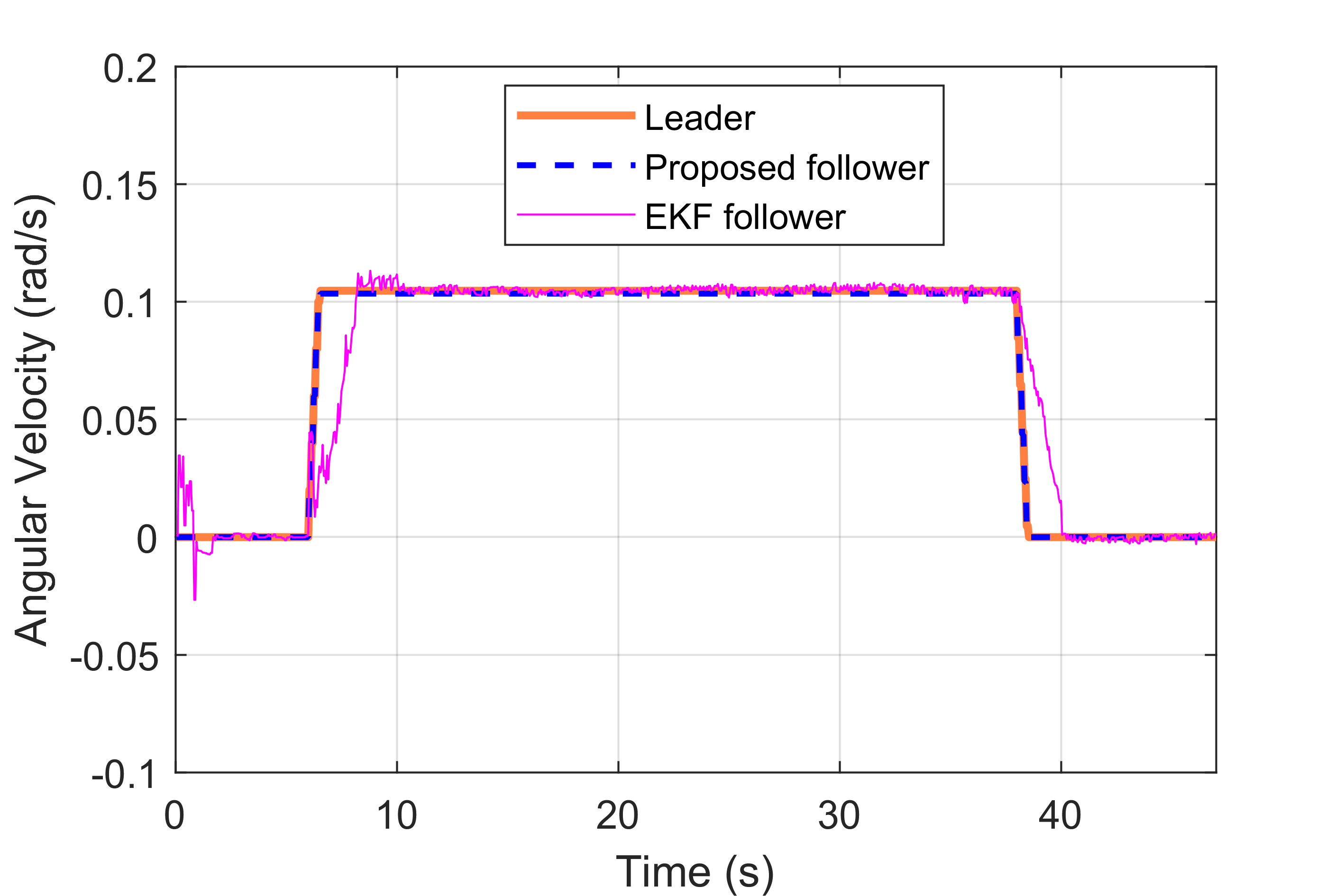}}
	\caption{Results of the third experiment. (a) Trajectory. (b)-(d) Formation errors. (e)-(f) Actual velocity, received velocity and estimated velocity of the leader.}
	\label{fig: results of the u-shaped path} 
	\vspace{-1.5em}
\end{figure}

In the third experiment, the leader is designed to move along a U-shaped trajectory consisting of two straight trajectory sectors and a semicircular sector, as shown in Fig. \ref{fig: results of the u-shaped path}(a), where the circles mark the starting points and the squares mark the ending points. The leader starts with a straight line at a velocity of $\mathbf{u}_l = [0.3, 0]^T$. When entering the semicircular sector, the leader changes its velocity to $\mathbf{u}_l = [0.1, \frac{\pi}{30}]^T$. After the semicircular sector is passed, the leader accelerates at a velocity of ${\mathbf{u}}_l = [0.3, 0]^T$ to complete the last straight sector. The moments when the leader crosses the two intersection points of the trajectory sectors are marked with black dotted lines in Figs. \ref{fig: results of the u-shaped path}(b)-\ref{fig: results of the u-shaped path}(d). The relative pose is initialized as $\mathbf{s}_0 = [0.9, 0.1, 0.31]^T$. The desired relative pose is set to $\overline{\mathbf{s}} = [0.6, 0, 0]^T$ during the linear motion, while it is switched to $\overline{\mathbf{s}} = [0.6, 0.15, \frac{\pi}{6}]^T$ during the circular motion. We also implemented the benchmark scheme using EKF velocity estimation for comparison purposes. Since both schemes reconstruct the leader's velocity, the trajectory switches and the associated velocity changes can be perceived by the follower. 

The experimental results of the third experiment are presented in Fig. \ref{fig: results of the u-shaped path}. From Fig. \ref{fig: results of the u-shaped path}(a), we can see that our follower achieves a smoother tracking trajectory when compared to the EKF-based scheme. Meanwhile, Figs. \ref{fig: results of the u-shaped path}(b)-\ref{fig: results of the u-shaped path}(d) show that the formation errors in our system converge faster and is stabler. In particular, the formation error $\varepsilon_x$ fluctuates much more gently when $\mathbf{u}_l$ changes at the intersections of the trajectory sectors. In Figs.~\ref{fig: results of the u-shaped path}(e)-\ref{fig: results of the u-shaped path}(f) we can observe that the velocity received by the proposed follower matches well with the leader's actual velocity, while the EKF estimation has considerable delay and jitter. Overall, the results indicate that the proposed leader-follower system is more responsive to velocity changes.

\section{Conclusions} \label{section: conclusions}
In this paper, we developed a ROS-compatible OISAC scheme that integrates camera sensing and SCC for cooperative mobile robotics. Our scheme addresses new problems such as image blur and long image display delays, and is designed for real-time control of mobile robots. Our experiments have validated the functionality of the proposed scheme. We focused on the leader-follower formation control as a case study, and designed an OISAC-augmented control system that enables the follower to use RGB images to estimate the relative pose to the leader and extract the state information sent by the leader. We implemented a new control law with proven stability and bounded errors to achieve accurate and stable formation control. Real-world experiments using two Turtlebot2 robots demonstrated the stability and robustness of the proposed scheme, and showed that the follower using the OISAC scheme and the devised control law is more responsive to the leader's movements than a benchmark system that uses EKF to estimate the leader's states. Future work includes trying other optical communication technologies (e.g., visible light communication) and adapting the OISAC scheme to more complex tasks, such as obstacle avoidance and cooperative object transportation.


\bibliographystyle{IEEEtran}
\bibliography{reference}

\end{document}